\documentclass[a4paper]{article}

\usepackage{INTERSPEECH2020}
\usepackage{amsfonts}
\usepackage{bm}
\usepackage{float}
\usepackage{graphicx}
\usepackage{algorithm}
\usepackage[noend]{algpseudocode}
\usepackage{mathtools}
\usepackage{hyperref}
\usepackage{tablefootnote}
\usepackage{xcolor}
\usepackage{xspace}

\def\espresso{\textsc{Espresso}\xspace}
\def\pychain{\textsc{PyChain}\xspace}
\def\pychainexample{\textsc{PyChain-example}\xspace}
\def\espnet{\textsc{ESPnet}\xspace}

\title{\pychain: A Fully Parallelized PyTorch Implementation of LF-MMI for End-to-End ASR}
\name{Yiwen Shao$^1$, Yiming Wang$^1$, Daniel Povey$^3$, Sanjeev Khudanpur$^{1,2}$\thanks{This work was partially supported by the DARPA GARD program and an unrestricted gift from \href{https://www.apptek.com/}{Applications Technology (AppTek)}.}}
\address{
  $^1$Center for Language and Speech Processing,
  $^2$Human Language Technology Center of Excellence,\\
  Johns Hopkins University, Baltimore, MD, USA \\
  $^3$Xiaomi Inc., Beijing, China
}
\email{\{yshao18, yiming.wang, khudanpur\}@jhu.edu, dpovey@gmail.com}

\begin{document}

\maketitle
\begin{abstract}
  We present \pychain, a fully parallelized PyTorch implementation of end-to-end lattice-free maximum mutual information (LF-MMI) training for the so-called \emph{chain models} in the Kaldi automatic speech recognition (ASR) toolkit. Unlike other PyTorch and Kaldi based ASR toolkits, \pychain is designed to be as flexible and light-weight as possible so that it can be easily plugged into new ASR projects, or other existing PyTorch-based ASR tools, as exemplified respectively by a new project \pychainexample, and \espresso, an existing end-to-end ASR toolkit. \pychain's efficiency and flexibility is demonstrated through such novel features as full GPU training on numerator/denominator graphs, and support for unequal length sequences.  Experiments on the WSJ dataset show that with simple neural networks and commonly used machine learning techniques, \pychain can achieve competitive results that are comparable to Kaldi and better than other end-to-end ASR systems.
\end{abstract}

\section{Introduction}

In the past few years, end-to-end or pure neural approaches to automatic speech recognition (ASR) have received a lot of attention. Among them, connectionist temporal classification (CTC) \cite{graves2013speech}, RNN-Transducer \cite{graves2012sequence} and sequence-to-sequence models with attention \cite{sutskever2014sequence, bahdanau2016end} are of high interests.
This trend is largely caused by two main reasons: (\romannumeral 1) the increasing demand on a simpler pipeline without several stages as in traditional hidden Markov model (HMM) based methods, and (\romannumeral 2) the easy access to the latest advances in deep learning, supported by powerful deep learning platforms like PyTorch \cite{NEURIPS2019_9015} and TensorFlow \cite{abadi2016tensorflow}. As a result, many end-to-end ASR toolkits have been developed and have achieved impressive results, such as Deep Speech \cite{amodei2016deep}, \espnet \cite{watanabe2018espnet} and \espresso  \cite{wang2019espresso}. 

On the other hand, in most scenarios (with small amount of data available especially) traditional hybrid (HMM-DNN) systems exemplified by Kaldi \cite{povey2011kaldi} perform better. In particular, chain models, with lattice-free MMI (LF-MMI) \cite{povey2016purely} training, are the state-of-the-art (SOTA) model in Kaldi \cite{povey2018semi}. Driven by the need for single stage training, an end-to-end version of LF-MMI (E2E LF-MMI) was proposed by Hadian et. al \cite{hadian2018end,hadian2018flat} which removes any dependencies on HMM-GMM
alignments, and the context-dependency trees customarily used in chain model training. But E2E LF-MMI is still implemented in Kaldi, and not compatible with PyTorch, TensorFlow etc.  

To bridge the gap between Kaldi and other mainstream deep learning platforms, a lot of excellent work has been done recently, such as PyTorch-Kaldi \cite{ravanelli2019pytorch}, PyKaldi \cite{pykaldi} and  PyKaldi2 \cite{lu2019pykaldi2}. However, none of them have managed to implement fully parallel LF-MMI training \footnote{PyKaldi2 only implemented LF-MMI loss with minibatch size 1, which is unable to achieve competitive results}, which tends to be the most effective and widely used loss function for training Kaldi ASR systems. 

\begin{figure}[t]
  \centering
  \includegraphics[trim={0 15mm 0 0}, clip, width=\linewidth, height=27mm]{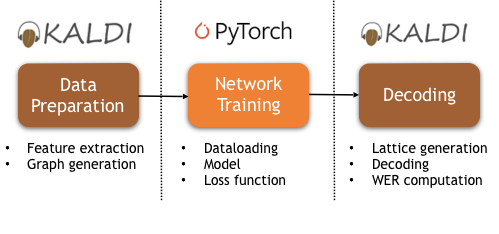}
  \caption{The pipeline of doing end-to-end LF-MMI training with \pychain.}
  \vspace*{-18pt}
  \label{fig:pychain_pipe}
\end{figure}

To fill in this gap, we present \pychain, a light-weight yet powerful PyTorch implementation of the E2E LF-MMI criterion written in C++/CUDA but wrapped with PyTorch.\footnote{\url{https://github.com/YiwenShaoStephen/pychain}} Also, we present examples of using \pychain ~in two different scenarios: (a) \pychainexample, a toy example written from scratch with only necessary utilities\footnote{\url{https://github.com/YiwenShaoStephen/pychain\_example}}, and (b) \espresso, a more integrated end-to-end ASR toolkit originally built for sequence-to-sequence models\footnote{\url{https://github.com/freewym/espresso}}. A high-level overview of the pipeline is shown in Figure~\ref{fig:pychain_pipe}. By doing so, we are also able to make a more direct comparison of the E2E LF-MMI method and other end-to-end approaches without using Kaldi-specific functionality, such as natural gradient SGD, and parameter averaging across parallel jobs \cite{povey2014parallel}.

We perform ASR benchmark experiments on WSJ dataset \cite{paul1992design}. Without bells and whistles, \pychain is able to achieve SOTA results comparable to Kaldi, and better than other end-to-end ASR systems, but using simpler models and techniques.

The rest of the paper is organized as follows. We briefly discuss the LF-MMI loss function in Section 2. And then in Section 3, we describe \pychain's architecture and implementation in detail. The experimental setup and results are shown in Section 4, and we present our conclusions in Section 5.

\section{End-to-End LF-MMI}
Maximum mutual information (MMI) \cite{povey2005discriminative}, is one of the most popular criteria for discriminative sequence training in ASR. It takes into account the entire word-sequence (utterance) holistically in the objective instead of only considering individual frames, as in frame-level functions. It aims to maximize the \emph{ratio} of the probability of the acoustics and reference transcription to that of the acoustics and all other possible transcriptions.

When MMI was first proposed, the marginal probability over all possible transcriptions was approximated using N-best lists, and later using lattices \cite{valtchev1996lattice, woodland2002large}. Subsequently, Povey et. al \cite{povey2016purely} proposed lattice-free MMI (LF-MMI) training by utilizing an n-gram phone language model for the \emph{denominator} computation. But they still require an alignment for the \emph{numerator} computation. Hadian et. al \cite{hadian2018end, hadian2018flat} extended LF-MMI training to not rely on any alignment, or even context-dependency trees, from a bootstrap model, such as an HMM-GMM system. In this work, we will mainly focus on this end-to-end (alignment-free) version of LF-MMI training.

\subsection{LF-MMI Loss}
For one utterance, MMI objective can be formulated as:
\begin{equation}
    F_{MMI}= \log\frac{P(\Vec{X}\rvert \Vec{W_{r}})P(\Vec{W_{r}})}{\sum_{\Vec{\hat{W}}}P(\Vec{X}\rvert \Vec{\hat{W}})P(\Vec{\hat{W})}}
    \label{MMI}
\end{equation}
where \Vec{X} is the input frames sequence, \Vec{W_{r}} is the gold transcription of \Vec{X}, and \Vec{\hat{W}} is any possible transcription. 

In LF-MMI \cite{povey2016purely}, a n-gram phone language model (LM) is integrated with the acoustic part to encode all possible word sequences into one single HMM graph called \textit{denominator} graph $\mathbb{G}_{den}$. Thus, we can replace the denominator part in Eq.~\eqref{MMI} with $P(\Vec{X}\rvert\mathbb{G}_{den})$.

Similarly, the numerator part in Eq.~\eqref{MMI} can be replaced with $P(\Vec{X}\rvert\mathbb{G}_{num})$ where $\mathbb{G}_{num}$ is the numerator graph generated by composing the true transcript to the denominator graph $\mathbb{G}_{den}$. We will use \textit{numerator} and \textit{denominator} for short in this paper. 

Extending to the corpus or batch level, we can get the final LF-MMI loss function as:
\begin{equation}
    F_{MMI}= \sum_{u=1}^U\log\frac{P(\Vec{X}^{(u)}\rvert \mathbb{G}_{num}^{(u)})}{P(\Vec{X}^{(u)}\rvert\mathbb{G}_{den}^{(u)})}
    \label{Graph}
\end{equation}
\subsection{LF-MMI Derivatives}
Another big advantage of MMI loss function is that, although seemly complicated, its derivatives can be finally reduced to the difference of occupation probability between numerator and denominator graphs[]:
\begin{equation}
    \frac{\partial F_{MMI}}{\partial \Vec{y}^{(u)}(s)} = \gamma_{num}^{(u)}(s) - \gamma_{den}^{(u)}(s)
    \label{deriv}
\end{equation}
where $\Vec{y}_{(u)}$ is the network output of the $u$-th utterance, which we interpret as the log-likelihood for each state. And $\gamma_{num}^{(u)}(s)$ and $\gamma_{den}^{(u)}(s)$ are the state occupation probability of state $s$ in the numerator and denominator graphs respectively, calculated by the Forward-Backward algorithm in HMM.

\section{\pychain}
\begin{figure}[t]
  \centering
  \includegraphics[width=\linewidth]{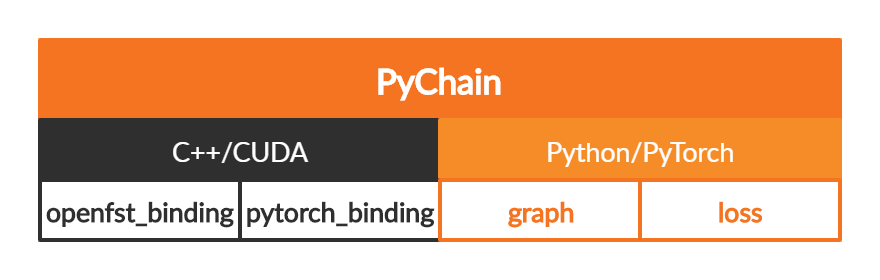}
  \caption{The layout of \pychain modules.}
  \vspace*{-12pt}
  \label{fig:pychain}
\end{figure}

Our work consists of two separate parts. The first part is \pychain, namely, the loss function itself written in C++/CUDA and wrapped in Python for PyTorch. It has following key components as shown in Figure~\ref{fig:pychain}:
\begin{itemize}
    \item \textit{openfst\_binding}: Functions to read Kaldi numerator/denominator graphs stored as Finite State Transducers (FSTs) format and transform them into tensors.
    \item \textit{pytorch\_binding}: Functions for Forward-Backward computation.
    \item \textit{graph.py}: Classes for HMM graphs (\texttt{ChainGraph} and \texttt{ChainGraphBatch}) 
    \item \textit{loss.py}: Loss function \texttt{ChainLoss} (\texttt{nn.Module}) and \texttt{ChainFunction} (\texttt{autograd.Function})
\end{itemize}

The second part is about the examples of using \pychain. In order to illustrate the easy-to-use property of \pychain, we give examples on either writing a speech recognition project from scratch (i.e. \pychainexample, which is similar to \texttt{PyTorch/examples}\footnote{\url{https://github.com/pytorch/examples}}), or plugging \pychain into an integrated ASR toolkit like \espresso.  It only takes to write basic PyTorch utilities such as dataloaders, models and train/test scripts with minimal codes.

\subsection{Pipeline}
As shown in Figure~\ref{fig:pychain_pipe}, we take advantage of both Kaldi and PyTorch to form a full recipe for an end-to-end ASR task. We do data preparations and final decoding in Kaldi for efficiency and consistency, but all other parts in PyTorch. Data preparation includes both feature extraction (e.g. MFCC) and numerator/denominator graph (FSTs) generation. Please note that, because there is no need for alignment in the E2E LF-MMI, we will not do any HMM-GMM pre-training in this stage. After all data is prepared, we go to PyTorch for dataloading and network training, which we will show in details below. After the model is trained and saved, we will load the data and model in PyTorch and then do a forward pass to get the output (posterior). They will be either piped to Kaldi for decoding on the fly, or dumped to the disk and then decoded afterwards.

\subsection{Input Data}
As in many other ASR toolkits \cite{watanabe2018espnet, wang2019espresso}, we use Kaldi for data preparation. Both acoustic features and numerator/denominator graphs (FSTs) will be saved in scp/ark format by Kaldi. We use kaldi\_io\footnote{\url{https://github.com/vesis84/kaldi-io-for-python}} to read matrix data like input features as numpy arrays and then transform them into PyTorch Tensors. For FSTs, we write our own functions in C++ based on \textit{OpenFST} \cite{allauzen2007openfst} and then bind it with Python using \textit{pybind} \cite{pybind11}, so that all these functions can be called in Python seamlessly.

\subsection{Numerator \& Denominator Graphs}
As shown in Eq.~\eqref{Graph}, HMM graphs are used as supervision in LF-MMI. We follow Kaldi's practice of using probability distribution function (pdf) to estimate the likelihood \cite{povey2011kaldi} of an HMM emission. As a result, both the network output and the occupation probability in Eq.~\eqref{deriv} will be computed with respect to a pdf-index (pdf-id) instead of an HMM state. And a typical transition in a HMM graph will be from a from-state $s_f$ to a to-state $s_t$ emitting a pdf-id $d$ with a probability $p$. In this way, a dense transition matrix of an HMM graph will be of size $(S, S, D)$ with transition probability $p$ in each cell, where $S$ is the total number of HMM states and $D$ is the total number of pdf-ids.

However, due to the heavy minimization and pruning done in LF-MMI, both numerator and denominator graphs are very sparse. In \pychain, we manage to represent numerator and denominator in a uniform way with a \texttt{ChainGraph} object $\mathbb{G}$ that is similar to the COO (Coordinate list) format for a sparse matrix:
\begin{itemize}
    \item A transition
        $\Vec{T}^{i} = \begin{bmatrix}s_{f}^{i}, &s_{t}^{i}, &d^{i}\end{bmatrix}$
    where $i$ is the index of the transition, $s_{f}$ is the from-state and $s_{t}$ is the to-state. $d$ is the pdf-id on this transition. It basically denotes the coordinate of a transition.
    \item The forward transition matrix
        $F_T = \begin{bmatrix} \Vec{T}^{i} \end{bmatrix}$ of size $(I, 3)$
    and forward transition probability vector
        $F_p = \begin{bmatrix} p^i \end{bmatrix}$ of length $I$ where $I$ is the total number of transitions in a graph. They act like the coordinate list and value vector for a COO sparse matrix.
    They are denoted as $F$ (meaning the "forward") as they are sorted in ascending order of the from-state $s_f$.
    \item The forward indices matrix 
        $F_I = \begin{bmatrix} (I_{start}^{s}, I_{end}^{s}) \end{bmatrix}$
    of size $(S, 2)$ where $I_{start}^{s}$ and $ I_{end}^{s}$ denote the start and end row index in $F_T$, between which are all transitions from $s_f$. 
\end{itemize}
Similarly, we have another three tensors for backward transitions, namely, $B_T, B_P$ and $B_I$. They are equivalent to these forward ones except that they are sorted by $s_t$. They are arranged in this way for a quick indexing of any transition by its from/to state.

In the end, there is a unique initial state $s_{init}$ for each graph and a final probability vector 
    $\Vec{p_{final}} = \begin{bmatrix} p^{s} \end{bmatrix}$ of length $S$
where $p^s$ is the probability of state $s$ being the final state in this HMM graph. 

As in most cases, LF-MMI will be trained with batches of utterance, we extend \texttt{ChainGraph} to its batch version, \texttt{ChainGraphBatch}. As its name suggests, it contains a batch of graphs and can be initialized by either a list of different ChainGraph objects (for numerators), or by a single \texttt{ChainGraph} object (for denominator). It has exact the same type of tensors as \texttt{ChainGraph} does, except that there is one more dimension for each of these tensors representing the batch dimension. And the default zero-padding is employed here when the sizes of these tensors differ.

\subsection{Forward-Backward Computations}
We follow the basic routine suggested by PyTorch custom C++ and CUDA extensions guide to write kernels for \texttt{ChainFunction}.

Our implementation has following key features:
\begin{itemize}
    \item Both numerator and denominator use the same piece of codes for forward-backward computation with full support for CPU/GPU computation.
    \item The computation is done in probability space instead of log-probability space by utilizing leaky HMM \cite{povey2016purely} to solve underflow issues for both numerator and denominator.
    \item It supports variable lengths of sequences without doing mandatory silence padding or speed perturbation \cite{ko2015audio} that was required by \cite{hadian2018end, hadian2018flat}.
\end{itemize}

A detailed description of our implementation is shown in Algorithm~\ref{code:forward}. We only present Forward algorithm as Backward algorithm is similar to the forward pass and we will not repeat here. Also, for simplicity and clarity, we omit the leaky HMM and readers are referred to \cite{povey2016purely} for details. 

In Algorithm~\ref{code:forward}, $L$ denotes the log-likelihood of pdf-ids along the sequences as the output from a neural network. It has size of $(B, T, D)$, where B is the batch size, T is the length and D is the total number of pdf-ids. We get a batch of variable lengths sequences by firstly sorting these sequences by their lengths in descending order and then do the zero-padding to the right. We use $B_{v}(t)$ to denote the valid batch size at each sequence step $t$ before any padding. For example, if there are 3 sequences of lengths $(100, 99, 98)$ in a batch, $B_{v}$ would be $[3, 3, \cdots, 2, 1]$. Forward trellis is represented by $\alpha$ of size $(B, T+1, S)$. Finally, we sum over the probability of each state in the final step of $\alpha$ to get the output $O$ of the forward algorithm, which is $\log P(\Vec{X}\rvert \mathbb{G})$ in Eq.~\eqref{Graph}.

\begin{algorithm}[htb]
\caption{The Forward Algorithm. Loops over sequences and states in line~\ref{code:start}-\ref{code:end} can be parallelized due to no dependency.}
\label{code:forward}
\hspace*{\algorithmicindent} \textbf{Input:} $L, B_v, B_T, B_p, B_I, p_{final}$. \\
\hspace*{1.5cm} network output $L$ of size (B, T, D); \\
\hspace*{1.5cm} valid batch size at each time step $B_v$ of length T;\\
\hspace*{1.5cm} backward transition matrix $B_T$ of size (B, I, 3);\\
\hspace*{1.5cm} backward transition probability $B_p$ of size (B, I);\\
\hspace*{1.5cm} backward transition index $B_I$ of size (B, S, 2);\\
\hspace*{1.5cm} final probability of each state $p_{final}$ of size (B, S); \\
\hspace*{\algorithmicindent} \textbf{Output:} $O$.\\
\hspace*{1.5cm} total log-probability $O$ of size (B); 
\begin{algorithmic}[1]
\Procedure{Forward}{$L, B_v, B_T, B_p, B_I, p_{final}$}
\State $\alpha[:, :, :] \coloneqq 0; \alpha[:, 0, s_{init}] \coloneqq 1$ \Comment{initialize $\alpha$}
 \For{$t\gets1$ \textbf{to} $T$}  \Comment{loop over time steps}
   \State $bs \coloneqq B_v[t]$
    \For{$b \gets0 $ \textbf{to} $bs-1$}  \Comment{loop over sequences} \label{code:start}
        \For{$s \gets 0$ \textbf{to} $S-1$} \Comment{loop over states} \label{code:end}
            \State $(I_{start}, I_{end}) \coloneqq B_{I}[b, s, :]$
            \For{$i \gets I_{start}$ \textbf{to} $I_{end} - 1$}
               \State $(s_{f}, s_{t}, d) \coloneqq B_{T}[b, i, :]$
                \State $p \coloneqq B_{p}[b, i]$
                \State $\alpha[b, t, s_t] \mathrel{{+}{=}} p\cdot\alpha[b, t-1, s_f]\cdot L[b, t, d]$
            \EndFor
         \If{t = T} \Comment{multiply final prob}
        \State $\alpha[b, t, s] \mathrel{{*}{=}} p_{final}[b, s]$
        \EndIf
        \EndFor
\EndFor
 \EndFor
 \State $O \coloneqq \alpha[:, T, :].\textrm{sum}(1)$ \Comment{sum over all states}
\EndProcedure
\end{algorithmic}
\end{algorithm}

\vspace{-1mm}
\section{Experiments}
We do most of our experiments on WSJ (Wall Street Journal) \cite{paul1992design} dataset, which is a database with 80 hours of transcribed newspaper speech. We consider the standard subsets \textit{si284}, \textit{eval92} for training and test, respectively. We use exactly the same subset from \textit{si284} for validation as in Kaldi, i.e., randomly selected 300 utterances plus their corresponding speed-perturbed versions (if there are any).

\subsection{Data Preparation}
The 40-dimensional MFCC extracted from 25 ms frames every 10 ms are used as input features. They are then normalized on a per-speaker basis to have zero mean. For numerator and denominator graphs generation, we adopt the best setting in \cite{hadian2018end, hadian2018flat}. The phone language model for the denominator graph is estimated using the training transcriptions (choosing a random pronunciation for words with alternative pronunciations in the phoneme-based setting), after
inserting silence phones with probability 0.2 between the words and with probability 0.8 at the beginning and end of the sentences. We use a trivial full biphone tree without CD (context-dependency) modeling and a 2-state-skip HMM topology.

\subsection{Model Architecture}
As one of our motivations for this work is to show the generality of the LF-MMI training outside Kaldi, we use common components to build our model, which only includes 1D dilated convolution (CNN or TDNN) \cite{peddinti2015time}, batch normalization \cite{ioffe2015batch}, ReLU \cite{nair2010rectified} and dropout \cite{srivastava2014dropout}. They are stacked 6 times in the sequence of \texttt{conv-BN-ReLU-dropout} with residual connections and finally followed by a fully connected layer. The network is of input dimension 40 (MFCC) and output dimension 84 (number of pdf-ids). The hidden dimension is 640 for all other layers.  The dropout rate is set to 0.2, and the convolutional layers are of kernel sizes of $(3, 3, 3, 3, 3, 3)$, strides of $(1, 1, 1, 1, 1, 3)$ (equivalent to a subsampling factor of 3 in the original LF-MMI), and dilations of $(1, 1, 1, 3, 3, 3)$.\footnote{The detailed configuration can been found in \url{https://github.com/freewym/espresso/blob/master/examples/asr_wsj/run_chain_e2e.sh}.}
\subsection{Training Schedule}
For training schedule, we use Adam \cite{kingma2014adam} as the optimization method. The learning rate is set to start from $10^{-3}$ and will be halved if no improvement on the validation set is seen at the end of each epoch, and finally fixed to $10^{-5}$. Similar to the findings in \cite{hadian2018flat}, we also find out that curriculum training \cite{bengio2009curriculum} (i.e. training utterance from short to long) is very helpful to E2E LF-MMI training in terms of robustness and ease to converge. However, we only do curriculum training for the first 1 or 2 epochs to achieve the best randomness in training. Otherwise, we sort all sequences in the corpus by its length so that sequences with similar lengths would form a minibatch. In other words, we only do shuffle on the batch level. Finally the model with the best validation loss is selected for decoding.

\subsection{Empirical Results}
We compare our results on WSJ \textit{eval92} with the original E2E LF-MMI and other SOTA end-to-end ASR systems. As shown in Table~\ref{tab:example}, we not only achieve competitive results but also use a smaller number of parameters and much simpler structure. Note that we only use n-gram LMs for decoding, while the others (except Hadian et al. \cite{hadian2018flat}) use more powerful neural LMs. 

Our implementation is also more efficient than the original one where extra time is wasted on the padded silence parts. Some might argue that since the Forward-Backward algorithm has a temporal dependency on its previous step, the longest sequence inside a batch will decide the computational time and thus our unequal length version will not make a difference on the efficiency. The explanation is, as shown in Algorithm~\ref{code:forward}, because we are doing parallel computation state-wise and sequence-wise, there would be at most $B \cdot S$ ($128 * 7,398 = 946, 944$ in our case) concurrent jobs on a GPU at once. However, there are much fewer CUDA cores in a modern GPU (e.g. 3584 for a NVIDIA GTX 1080 Ti in this experiment) which becomes the actual bottleneck for speed. As a result, the computational time will be proportional to the number of jobs we have in total. 

Flexibility is another big advantages of our variable lengths supported implementation. Instead of modifying utterance lengths by either silence padding or speed perturbation, we are able to form any utterances into a minibatch to get the maximal flexibility and randomness in training. Note that simply doing zero-padding at the end of each utterance does not work easily because these padded features may not be necessarily aligned to the final states and would probably lead to a terrible alignment and thus degrade the results.

\begin{table}[htb]
  \caption{WERs (\%) on the WSJ without data augmentation.}
  \label{tab:example}
  \centering
  \begin{tabular}{l c c}
    \toprule
    \textbf{System} & \textbf{\# Params (M)} &  \textbf{WER (\%)} \\
    \midrule
    Zeghidour et al. \cite{zeghidour2018end} & 17&5.6 \\
   Baskar et al. \cite{baskar2019promising} &$\sim$100 &3.8 \\
   Likhomanenko et al. \cite{likhomanenko2019needs} &17 & 3.6 \\
   Zeghidour et al. \cite{zeghidour2018fully} &17 & 3.5 \\
   Wang et al. \cite{wang2019espresso} & 18 & \textbf{3.4} \\
   \midrule
   Hadian et al. \cite{hadian2018flat} & 9.1 & 4.3 \\ 
   \pychain   & 6.3 & 3.5 \\
    \bottomrule
  \end{tabular}
\end{table}

For the completeness of comparison, we also do experiments on augmented data with a 2-fold speed perturbation (sp) as was originally required by \cite{hadian2018end} and compare the results with other end-to-end systems with data augmentation. As shown in Table~\ref{tab:example2}, we again match up to the results in Kaldi but with a smaller network.

\begin{table}[htb]
  \caption{WERs (\%) on the WSJ with data augmentation.}
  \label{tab:example2}
  \centering
  \begin{tabular}{l c c}
    \toprule
    \textbf{System} & \textbf{\# Params (M)} &  \textbf{WER (\%)} \\
    \midrule
   An et al. \cite{an2019cat} & 16 & 3.2 \\
   Amodei et al.\tablefootnote{12k hours AM train set and common crawl LM.} \cite{amodei2016deep} & - &\it\textcolor{gray}{3.1} \\
   Kriman et al.\tablefootnote{Data augmentation and pre-trained on LibriSpeech \cite{panayotovcpk15} and \href{https://voice.mozilla.org/en/datasets}{Mozillas Common Voice datasets}.} \cite{kriman2019quartznet} & 19 &\it\textcolor{gray}{3.0} \\
   \midrule
   Hadian et. al \cite{hadian2018flat} & 9.1 & \textbf{3.0} \\ 
   \pychain   & 6.3 & \textbf{3.0}  \\
    \bottomrule
  \end{tabular}
\end{table}


\vspace{-2mm}
\section{Conclusions and Future Work}
We presented \pychain, a PyTorch-based implementation of the end-to-end (alignment-free) LF-MMI training method for ASR.  We used tensors to represent HMM graphs, which permits seamless transformation between Kaldi and PyTorch. Examples in \pychainexample and \espresso ~illustrate that \pychain ~can be easily used in a project that includes building an ASR system from scratch, or extending an existing ASR system.  We would like to support the use of \pychain in such projects in the future. 

Experiments on WSJ with simple networks exhibit the power of \pychain. It is expected that, with larger neural networks and more sophisticated methods, \pychain has the potential to obtain even better results on many other ASR tasks.

In the future, we plan to extend \pychain to support regular LF-MMI and further bridge the gap between Kaldi and PyTorch. We hope that our experience with \pychain will inspire other efforts to build next-generation hybrid ASR tools.

\bibliographystyle{IEEEtran}
\bibliography{mybib}


\end{document}